\documentclass[aps,pra,reprint,superscriptaddress,showpacs]{revtex4-1}
\pdfoutput=1

\usepackage[cp1250]{inputenc}
\usepackage{amsmath}
\usepackage{amsfonts}
\usepackage{textcomp}
\usepackage{graphicx}
\usepackage{epstopdf}
\usepackage{mathrsfs}
\usepackage{amsbsy}
\usepackage[rightcaption]{sidecap}
\usepackage{hyperref} 
\usepackage{cleveref}
\usepackage{xr}   

\DeclareMathOperator{\sn}{sn}
\DeclareMathOperator{\cn}{cn}
\DeclareMathOperator{\dn}{dn}

\DeclareMathOperator{\di}{d\kern-0.4ex}

\hypersetup{pdfauthor={Wiktor Walasik,  Gilles Renversez},pdftitle={Plasmon--soliton waves in planar slot waveguides: I. Modeling}}

\begin{document}

\title{Plasmon--soliton waves in planar slot waveguides: I. Modeling}

\author{Wiktor Walasik}
\affiliation{Aix--Marseille Universit\'{e}, CNRS, Centrale Marseille, Institut Fresnel, UMR 7249, 13013 Marseille, France}
\email[]{gilles.renversez@fresnel.fr}
\affiliation{ICFO --- Institut de Ci\`{e}ncies Fot\`{o}niques, Universitat Polit\`{e}cnica de
Catalunya, 08860 Castelldefels (Barcelona), Spain}
\author{Gilles Renversez}
\affiliation{Aix--Marseille Universit\'{e}, CNRS, Centrale Marseille, Institut Fresnel, UMR 7249, 13013 Marseille, France}

\date{\today}

\begin{abstract}
We present two complementary models  to study  stationary nonlinear solutions in one-dimensional  plasmonic slot waveguides made of a finite-thickness nonlinear dielectric core surrounded by metal regions. The considered nonlinearity is of focusing Kerr type. 
In the first model, it is assumed that the nonlinear term depends only on the transverse component of the electric field and that the nonlinear refractive index change is small compared to the linear part of the refractive index. 
This first model allows us to describe analytically the field profiles in the whole waveguide using Jacobi elliptic special functions. It also provides a closed analytical formula for the nonlinear dispersion relation. 
In the second model, the full dependency of the Kerr nonlinearity on the electric field components is taken into account and no assumption is required on the amplitude of the nonlinear term. The disadvantage of this approach is that the field profiles must be computed numerically. Nevertheless analytical constraints are obtained to reduce the parameter space where the solutions of the nonlinear dispersion relations are sought.
\end{abstract}


\maketitle

\section{Introduction}

Studies of stationary nonlinear waves possessing the properties of both plasmons and solitons started in the early 80s when this type of waves was described by Agranovich \textit{et al.}~\cite{Agranovich80}. Two main types of structures supporting plasmon--soliton waves  were studied. The first type contains one or two semi-infinite nonlinear media and was extensively studied in Refs.~\cite{Maraduduin83,Stegeman84,Stegeman84c,Stegeman85,Ariyasu85,Mihalache86,Boardman87, Mihalache87, Yin09,Bliokh09,Walasik12,Huang12,Milian12,Ferrando13,Walasik14}. Transverse electric (TE) and transverse magnetic (TM) polarized waves were investigated in configurations built of semi-infinite nonlinear layers, a metal layer, and, in some cases, additional dielectric layers.
The second type of structures contains a finite-size nonlinear dielectric layer sandwiched between two semi-infinite metal layers. This type of structures will be called here nonlinear slot waveguide (NSW). Studies of structure with a nonlinear dielectric core started in the early 80s from fully dielectric structures~\cite{Fedyanin82,Holland86,Boardman86a,Chen88, Langbein83, Akhmediev89b,Mihalache94b}. The solutions of Maxwell's equations in a finite size nonlinear dielectric were given in terms of Jacobi elliptic functions~\cite{Abramowitz72}. A symmetry breaking bifurcation was predicted for the fundamental symmetric mode giving birth to an asymmetric mode~\cite{Holland86,Boardman86a}. Various methods to study fully dielectric waveguides with both nonlinear core and cladding were proposed~\cite{Langbein_85b,Sammut91,Chiang93,Li91,Sammut93,Chiang94,Capobianco95,Stathopoulos97}.

In 2007, Feigenbaum and Orenstein~\cite{Feigenbaum07} made the first attempt to study waveguides with a nonlinear core surrounded by a metal cladding (NSWs) instead of a dielectric one. Their method describes sub-wavelength confinement of light in two-dimensional plasmon--soliton beams propagating in NSWs. Such a strong confinement is ensured by a linear plasmon profile in the transverse direction and by the self-focusing effect in the lateral direction.
Recently, numerical~\cite{Davoyan08} and semi-analytical~\cite{Rukhlenko11,Rukhlenko11a} methods have been developed to study up to the three first nonlinear modes in NSWs.   Higher order modes were also reported in NSWs~\cite{Walasik14a}.

The NSW is an interesting and promising configuration for two reasons: (i) from the practical point of view, it is easier to fabricate high quality thin nonlinear films than bulky layers, like those needed in the configurations with semi-infinite nonlinear medium and  (ii) it has numerous potential applications. Devices based on the NSW configuration can be used for phase matching in higher harmonic generation processes~\cite{Davoyan09a} and for nonlinear plasmonic couplers~\cite{Salgueiro10,Davoyan11a}. Nonlinear switching~\cite{Nozhat12} was theoretically predicted in NSW-based structures that is similar to the nonlinear switching in graphene couplers~\cite{Smirnova13}. Tapered NSWs might be used for nanofocusing and loss compensation in order to enhance nonlinear effects~\cite{Davoyan10}.

The field of NSWs is relatively young and there is not a lot of works describing the properties of these structures. In this article, we build two new and complementary models that allow us to study efficiently and accurately the NSW configurations. These models have been briefly introduced in Ref.~\cite{Walasik14a} and here we present their detailed derivations. 
 
In Sec.~\ref{sec:problem}, a general description of the problem studied here is given. Section~\ref{sec:models} presents the theoretical derivation of our two models. In Sec.~\ref{sec:ver}, the validity of the model formulations is verified in the limiting case of a semi-infinite medium. 
These models are used to thoroughly study the propagation of plasmon-soliton waves in NSWs in the following article~\cite{Walasik14c} where analytical and numerical stability analysis results are also provided.

\section{Problem statement}
\label{sec:problem}

\begin{figure}[!b]
\centering
\includegraphics[width=0.95\columnwidth,clip=true,trim= 0 0 0 0]{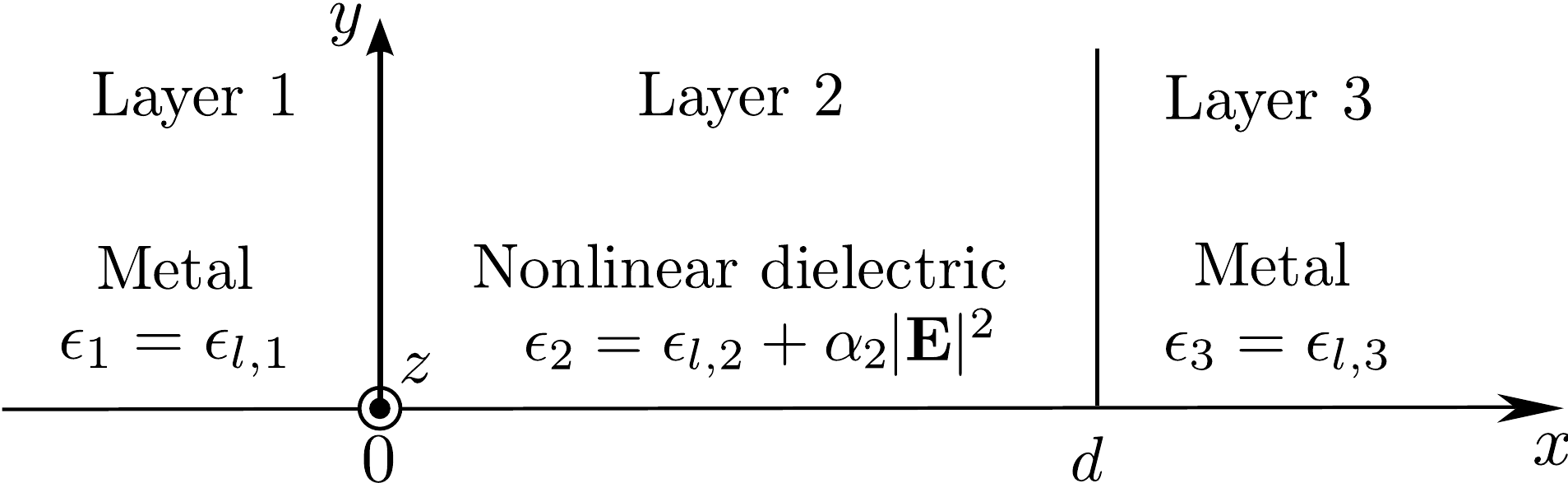}
\caption{Geometry of the plasmonic NSW with the parameters of the structure.}
\label{fig:geometry-slot}
\end{figure}

This article presents a derivation of the dispersion relations for the stationary TM polarized waves propagating in one-dimensional NSWs depicted schematically in Fig.~\ref{fig:geometry-slot}. 
We propose two models to study the light propagation in such structures. The first model is based on the approach proposed for fully dielectric structures in~\cite{Chen88,Fedyanin82} but is extended to structures containing metals. Due to the presence of the metal cladding, the field continuity conditions result in a new type of field profiles and very rich dispersion relations. This approach uses an approximated treatment of the nonlinearity in the Kerr medium which allows us to write and solve a single nonlinear wave equation in the finite-size nonlinear medium. Using the field continuity conditions at the core interfaces located at $x=0$ and $x=d$ the analytical formulas for the dispersion relations and for the field profiles are obtained in terms of Jacobi elliptic functions~\cite{Abramowitz72}. Therefore, this model will be called Jacobi elliptic function based model (JEM).

The second model is based on the approaches from Refs.~\cite{Mihalache87,Yin09} for a single interface between a nonlinear dielectric and a metal, and therefore it is named the interface model (IM). This model uses a more realistic treatment of the nonlinear Kerr effect than the JEM (\textit{i.e.} Kerr nonlinearity depends on all the components of the electric field). It allows us to obtain separate dispersion relations on the two interfaces of the NSW in analytical forms.
Comparing the dispersion equations for the left interface and for the right interface, results in an analytical condition that reduces the parameter space in which the solutions of Maxwell's equations in NSWs are sought.
The solutions are found by the numerical integration of Maxwell's equation in the core which allows then to relate the two interfaces. Maxwell's equations in the core are solved using the shooting method~\cite{Press07}. If the result of integration is consistent with the previously assumed values of the field and its derivatives at the slot interfaces then the corresponding $\beta$ is accepted as a genuine solution in our problem. 

The stationary solutions in our one-dimensional geometry are sought in the form of monochromatic harmonic waves:
\begin{equation}
\begin{Bmatrix}
\pmb{\mathscr{E}}(x,z,t)\\ 
\pmb{\mathscr{H}}(x,z,t)
\end{Bmatrix} = 
\begin{Bmatrix}
\textbf{E}(x)\\ 
\textbf{H}(x)
\end{Bmatrix} e^{i( k_0 \beta z - \omega t)},
\label{eqn:harmonic1}
\end{equation}
where $\textbf{E} = [E_x,0,iE_z]$ and $\textbf{H} = [0,H_y,0]$. We write the imaginary unit $i$ in front of the $z$-component of the electric field so that all the quantities $E_x, E_z$, and $H_y$ are real.
The propagation direction is chosen to be $z$ and $\omega$ denotes the angular frequency of the wave. $k_0 = \omega / c$ denotes the wavenumber in vacuum, $c$ denotes the speed of light in vacuum, and $\beta$ denotes the effective index of the propagating wave. The structure is invariant along the $y$ direction and therefore, it is assumed that the field profiles are invariant along the $y$ coordinate.

\section{Model derivation}
\label{sec:models}

\subsection{Maxwell's equations}

The derivation of our models starts from the general form of the Maxwell's equations in case of nonmagnetic materials (relative permeability~$\mu = 1$) without free charges ($\rho_f = 0$) and free currents ($\textbf{J}_f = 0$)~\cite{Jackson99}. For the harmonic monochromatic TM waves described by Eq.~(\ref{eqn:harmonic1}) and for an isotropic relative permittivity $\epsilon$ the Maxwell's equations read

\begin{subequations}
\label{eqn:max4}
\begin{align}
k_0 \beta E_x - \frac{dE_z}{d x} &= \omega \mu_0 H_y,
\label{eqn:rotE4}\\
E_x &= \frac{\beta}{\epsilon_0 \epsilon c} H_y,
\label{eqn:rotH41}\\
E_z &= \frac{1}{\epsilon_0 \epsilon \omega} \frac{d H_y}{d x}.
\label{eqn:rotH42}
\end{align} 
\end{subequations}

The nonlinearity studied here is of the Kerr type 
\begin{equation}
\epsilon = \epsilon_{l}(x) + \epsilon_{\textrm{nl}}(x),
\label{eqn:perm_nonl}
\end{equation} 
where $\epsilon_{l}$ denotes the linear, real part of the permittivity, $\epsilon_{\textrm{nl}} = \alpha(x)|\textbf{E}(x)|^2$ denotes the nonlinear part of the permittivity limited to the isotropic optical Kerr effect that depends on the electric field intensity, and $\alpha(x)$ denotes the function that takes values of the nonlinear parameters associated with different layers (in linear materials it is null). The imaginary part of the permittivity is neglected in the modal studies presented here. 

The formulation of the Kerr effect used in the following of this study can be found in the majority of the works on the nonlinear waveguides and nonlinear surface waves~\cite{Tomlinson80,Maradudin81,Leung85,
	Mihalache87,Lederer_83,Robbins83,Stegeman84II,Lederer_83II,Mihalache85, Seaton85,Seaton84,Langbein_85,Langbein_85II,Akhmediev82,Moloney86,Chiang93, Chiang94,Akhmediev89a,Micallef98,Sammut92,Chelkowski87,Torres94,Leine88,Mihalache94,Mihalache94a,
	Mihalache89,Holland86,Boardman86a,Chen88,Fedyanin82,
	Langbein83,Akhmediev89b,Mihalache94b,Li91,Sammut91,Langbein_85b,Fontaine90,Sammut93, Agranovich80,Stegeman85, Maraduduin83,Bliokh09,Davoyan09,Stegeman84,Stegeman84c, Ariyasu85,Mihalache86,Yin09,Huang12,Walasik12,Milian12,Ferrando13,Walasik14,Degiron10,Huang12a,
	Feigenbaum07,Davoyan08,Rukhlenko11,Rukhlenko11a,Huang09}. It describes sufficiently well the nonlinear effects studied here. To describe the physics of our system we do not need to use the more complex form of the Kerr nonlinear term which was described in Refs.~\cite{Crosignani82,Ciattoni05}.

\subsection{Jacobi elliptic function based model}
\label{sec:JEM-theory}

\subsubsection{Nonlinear field profiles}
\label{sec:JEM-NL-field-prof}

We start the presentation of the models for NSWs with the approach that uses strong assumptions on the form of the nonlinear Kerr term but it provides the dispersion relations and the field profiles in analytical forms. This model provides more insight and understanding of the nature of the problem of finding stationary solutions in NSWs than the second, more numerical model. First, we will solve the nonlinear wave equation inside the waveguide core and find the nonlinear field profiles. Knowing the field profiles, we will be able to derive the dispersion relations for the NSW using the continuity conditions on the two nonlinear core interfaces.

In the frame of the Jacobi elliptic function based model (JEM), the Kerr-type nonlinearity is not treated in an exact manner. We assume that the nonlinear response of the material is isotropic and depends only on the transverse component of the electric field $E_x$ in the following way~\cite{Walasik12,Walasik14,Walasik14a}:
\begin{equation}
\epsilon(x) = \epsilon_l(x) + \alpha(x) E_x^2(x).
\label{eqn:kerr_simple-jem}
\end{equation} 
Functions $\epsilon_l(x)$ and $\alpha(x)$ are step-wise functions which take the values indicated in Table \ref{tab:eps-jem} depending on the layer (see Fig.~\ref{fig:geometry-slot} for layer number).
\renewcommand{\arraystretch}{1.2}
\begin{table}[ht]
\centering
  \begin{tabular}{ l  c  c  c }
    \hline
    \hline
    Layer \;\;& \;\;\;\;\;\;\;\;Abscissa\;\;\;\;\;\; \;\;& $\epsilon_l(x)$   &\;\;\;\;\;\;\;\;\;\; $\alpha(x)$ \;\;\;\;\;\;\;\;\;\; \\ \hline
    1 & $x<0$             &  $\epsilon_1 = \epsilon_{l,1}$      & 0 \\ 
    2 & $0\le x \le d $     &  $\epsilon_{l,2}$              & $\epsilon_0 c \epsilon_{l,2} n_2^{(2)} = \alpha_2$ \\
    3 & $   x > d $ &  $\epsilon_3 = \epsilon_{l,3}$            & 0 \\
    \hline
   \hline
  \end{tabular}
\caption{Values of the functions $\epsilon_l(x)$ and $\alpha(x)$ describing the properties of the materials in different layers. The nonlinear parameter $n_2$ in layer 2 is denoted by $n_2^{(2)}$.}
  \label{tab:eps-jem}
\end{table}

The derivation of the JEM starts from Maxwell's equations [Eqs.~(\ref{eqn:max4})]. These equations are combined together with Eq.~(\ref{eqn:kerr_simple-jem}) and with the help of the approximations about small nonlinear permittivity change we obtain the nonlinear wave equation:
\begin{equation}
\frac{\mathrm d^2 H_y}{\di x^2} - k_0^2 q^2(x) H_y + k_0^2 a(x) H_y^3 = 0,
\label{eqn:wave5}
\end{equation} 
where 
\begin{align}
q^2(x) &= {\beta^2 - \epsilon_l(x)},
\label{eqn:q-def}\\
a(x) &={\beta^2 \alpha(x)}/{\left[\epsilon_0 \epsilon_l(x) c\right]^2}.
\label{eqn:a-def}
\end{align}
A detailed derivation of Eq.~(\ref{eqn:wave5}) is presented in Refs.~\cite{Walasik14, Walasik-thesis14}.
We use the first integral treatment approach~\cite{Yin09,Mihalache87} and integrate Eq.~(\ref{eqn:wave5}) with respect to $x$. The result reads
\begin{equation}
\left(\frac{\di  H_y}{\di x}\right)^2 - k_0^2 q(x)^2 H_y^2 + k_0^2 \frac{a(x)}{2} H_y^4 = c_0.
\label{eqn:first_int}
\end{equation} 
The left-hand side of this equation gives us a formula for a quantity that is conserved along the transverse profile of the core of our one-dimensional nonlinear waveguide. Regardless of at which position $x$ we calculate it, the result will always be equal to the integration constant $c_0$. For the structures with a semi-infinite nonlinear medium~\cite{Walasik14}, the integration constant was set to zero, because both the magnetic field $H_y$ and its derivative $\mathrm d H_y/\mathrm d x$ tend to zero as $ x \rightarrow \pm \infty$.
Therefore, in semi-infinite cladding layers, we must set the integration constant $c_0 = 0$. Additionally, in these linear layers $a(x)$ is equal to zero. Thus, in the cladding, Eq.~(\ref{eqn:wave5}) reduces to a standard linear wave equation whose solutions are given by:
\begin{subequations} 
\label{eqn:met_fields}
\begin{align}
 H_1 &= H_0 e^{k_0 q_1 x}         \;\;\;\; \;\;\;\ \;\;\;\; \;\;\;\;  \;\;\;\; \;\;\;\;  \textrm{for }  -\infty \le x < 0, \label{eqn:met_fields_a}\\
 H_3 &= H_d e^{-k_0 q_3 (x - d)} \;\;\;\; \;\;\;\ \;\;\;\; \;\;\;\;   \textrm{for }   d \le x < +\infty,
\end{align}
\end{subequations} 
where only the appropriate exponential solutions are considered. Here the magnetic field amplitudes at the interfaces $x=0$ and $x=d$ are denoted by $H_0$ and $H_d$, respectively and $q_k$ denotes a constant value of the $q(x)$ function in $k$-th layer (for $k \in \{1, 2 ,3\}$).
As $H_y$ is the only component of the magnetic field, in the following derivation we omit the $y$ subscript and instead we use a subscript that enumerates the layer in which the field profile is defined (see Fig.~\ref{fig:geometry-slot}). 
 
The integration constant $c_0$ can be expressed as a function of the magnetic field amplitude $H_0$ 
using the continuity conditions for the tangential electromagnetic field components ($H_y$, $E_z$) at $x=0$ in Eq.~(\ref{eqn:first_int}): 
\begin{equation}
c_0 = k_0^2 \left[ \left(  {\epsilon_{l,2}}/{\epsilon_1}\right)^2 q_1^2 - q_2^2 + {a_2} H_0^2/2 \right] H_0^2,
\label{eqn:c0}
\end{equation} 
where $a_2$ denotes a constant value of $a(x)$ function in the nonlinear core (layer 2) and where, based on the assumption that the nonlinear permittivity change is small, we have substituted $\epsilon_2|_{x=0^+}$ by $\epsilon_{l,2}$ in the fraction numerator. Similar expression can be obtained for the second interface $x=d$.
Equations~(\ref{eqn:first_int}) and (\ref{eqn:c0}) allow us to find the sign of the integration constant $c_0$ for certain types of solutions. The solutions can be later classified according to the sign of  $c_0$.
Looking at Eq.~(\ref{eqn:first_int}) we notice that for $H_y$ field profiles that cross zero, at the point where $H_y = 0$, the only nonzero term on the left-hand side of this equation is $(\mathrm d H_y/\mathrm d x)^2$, which is strictly positive. Therefore, for this type of solutions, $c_0$ can only be positive. 
From Eq.~(\ref{eqn:c0}) we notice that for negative $q_2^2$ the value of the integration constant $c_0$ is also positive.

We are searching for guided waves in three-layer structures. Looking at Eqs.~(\ref{eqn:met_fields}), we notice that the condition for the waves to be localized in the waveguide core is satisfied when both $q_1$ and $q_3$ are real and positive quantities. In order to satisfy this condition, we will look only for the solutions with $\beta > \max\{\epsilon_1, \epsilon_3\}$ [see the definition of $q(x)$ and $q_k$ given by Eq.~(\ref{eqn:q-def})]. The quantity $q_2$ can be either real or imaginary leading to positive or negative values of $q_2^2$. 

In order to find the solutions of the nonlinear wave equation [Eq.~(\ref{eqn:wave5})] in the nonlinear core, we rewrite its first integral [Eq.~(\ref{eqn:first_int})] in the form:
\begin{equation}
\frac{\di H_2}{\sqrt{c_0 + k_0^2 q_2^2 H_2^2 - k_0^2 \frac{a_2}{2}H_2^4}} = \pm \di x,
\label{eqn:first_2}
\end{equation} 
which is then transformed to 
\begin{equation}
\frac{\di H_2}{\sqrt{A c_0 + A Q H_2^2 - H_2^4}} = \pm\; \sqrt{\frac{1}{A}}\; \di x,
\label{eqn:first_3}
\end{equation}  
using the reduced parameters $Q$ and $A$:
\begin{subequations}
\label{eqn:def-AQ-finite}
\begin{align}
Q &= k_0^2 q_2^2, \label{eqn:def-Q-finite}\\
A &= (k_0^2 a_2/2)^{-1}. \label{eqn:def-A-finite}
\end{align}
\end{subequations}

In this study, we deal only with the focusing Kerr-type dielectrics, therefore $A$ is always positive. Parameter $Q$ can be either positive or negative depending on the sign of $q_2^2$. 

Solutions of Eq.~(\ref{eqn:first_3}) take different forms depending on the sign of parameters $c_0$. We will solve this equation in two cases depending on the sign of $c_0$. The details of this derivation are provided in Appendix~\ref{app:jem}.

\textbf{I. The case $c_0>0$}

At first, we consider the case where $c_0>0$. The solution of the nonlinear wave equation [Eq.~(\ref{eqn:first_int})] for this case is given by 
\begin{align}
H_2(x) &=  \delta \cn\left\{   \sqrt{{{s}/{A}}}\; (x-x_0)  \big| m \right\},
\label{eqn:elip6} 
\end{align} 
where $\cn(u|m)$ is a Jacobi elliptic function with an argument $u$ and a parameter $m$~\cite{Abramowitz72}, and 
\begin{subequations} 
\label{eqn:gam_del1a}
\begin{align}
m &= {\delta^2}/{s}, \\
s &= \gamma^2 + \delta^2, \\
\gamma^2 &= ({ \sqrt{A^2Q^2 + 4 A c_0} - AQ})/{2}, \label{eqn:gam2} \\
\delta^2 & =  ({ \sqrt{A^2Q^2 + 4 A c_0} + AQ})/{2},  \mathrm{and} \label{eqn:del2}\\
x_0 &= -\sqrt{{{A}/{s}}} \cn^{-1}\left[{H_2(0)}/{\delta} \big| m\right].
\label{eqn:x01}
\end{align}
\end{subequations} 
Here $\cn^{-1}(u|m)$ is the inverse of the Jacobi elliptic function $\cn(u|m)$.

\textbf{II The case $c_0<0$}

For $c_0<0$, the solution of the nonlinear wave equation [Eq.~(\ref{eqn:first_int})] is given by 
\begin{align}
H_2(x) &=  \gamma \dn\left\{ \sqrt{{{\gamma^2}/{A}}}\; (x -x_0)   \big| m \right\},
\label{eqn:elip16f}
\end{align}
where $\dn(u|m)$ is a Jacobi elliptic function~\cite{Abramowitz72} and 
\begin{subequations} 
\label{eqn:gam_del2}
\begin{align}
m &= ({\gamma^2 - \delta^2})/{\gamma^2}, \label{eqn:param2}\\
\gamma^2 &= ({ AQ + \sqrt{A^2Q^2 - 4 A |c_0|}})/{2}, \\
\delta^2 &=  ({ AQ - \sqrt{A^2Q^2 - 4 A |c_0|}})/{2},\\
x_0 &= - \sqrt{{{A}/{\gamma^2}}}\dn^{-1}\left[{H_2(0)}/{ \gamma} \big| m\right]. \label{eqn:x02}
\end{align}
\end{subequations} 
Here $\dn^{-1}(u|m)$ is the inverse of the Jacobi elliptic function $\dn(u|m)$. Knowing the analytical form of the solutions in each layer of the structure [Eqs.~\ref{eqn:met_fields}, and (\ref{eqn:elip6}) or (\ref{eqn:elip16f})] we can derive the nonlinear dispersion relations for the cases of positive and negative $c_0$ values.

\subsubsection{Nonlinear dispersion relations}
\label{sec:disp-rel-slot}

\textbf{I. The case $c_0>0$}

Firstly, we derive the dispersion relation for NSWs when $c_0>0$.
Using the analytical formula for the field profile in the nonlinear core [Eq.~(\ref{eqn:elip6})], the field profiles in metal claddings given by Eqs.~(\ref{eqn:met_fields}), and Maxwell's equations [Eqs.~(\ref{eqn:max4})], we write the continuity conditions for the tangential electromagnetic field components $H_y$ and $E_z$ at the interfaces between the nonlinear core and the metal cladding $x=d$ \{the continuity conditions at $x=0$ was already used to evaluate the integration constant [see Eq.~(\ref{eqn:c0})]\}
\begin{enumerate}
\item  The continuity condition for the magnetic field component
\begin{equation}
H_2|_{x=d^-} = H_3|_{x=d^+}
\end{equation} 
yields
\begin{equation}
 \delta \cn\left[ \sqrt{{{s}/{A}}}\; (d-x_0) \;   \big|\; m \right] = H_d.
\label{eqn:hd-1}
\end{equation} 

\item The continuity condition for the tangential electric field component
\begin{equation}
E_{z,2}|_{x=d^-} = E_{z,3}|_{x=d^+}
\end{equation} 
transformed with the use of Eq.~(\ref{eqn:rotH42}), the formulas for the Jacobi elliptic function derivatives, and the symmetry properties of Jacobi elliptic functions gives
\begin{align}
{\delta \epsilon_3 \sqrt{s}}/({k_0 q_3 \epsilon_{l,2}} \sqrt{{A}})
\sn\left[ \sqrt{{{s}/{A}}}\; (d-x_0) \;   \big|\; m \right]  \nonumber  \\ 
\dn\left[ \sqrt{{{s}/{A}}}\; (d-x_0) \;   \big|\; m \right] = H_d,  
  \label{eqn:hd-2}
\end{align} 
where, based on the assumption that the nonlinear permittivity change is small, we substituted $\epsilon_2|_{x=0^+}$ by $\epsilon_{l,2}$ in the denominator on the left-hand side.
\end{enumerate}
Comparing the two expressions for $H_d$ given by Eqs.~(\ref{eqn:hd-1}) and (\ref{eqn:hd-2}), we obtain the nonlinear dispersion relation in its final form for the case of $c_0>0$
\begin{widetext}
\begin{align}
{k_0 q_3 \epsilon_{l,2}}\sqrt{A}\cn\left[ \sqrt{{{s}/{A}}}\; (d-x_0) \;   \big|\; m \right] = 
{ \epsilon_3}{k_0 q_3 \epsilon_{l,2}} \sqrt{{s}}
\sn\left[ \sqrt{{{s}/{A}}}\; (d-x_0) \;   \big|\; m \right]
\dn\left[ \sqrt{{{s}/{A}}}\; (d-x_0) \;   \big|\; m \right].
\label{eqn:disp-JEM-1a}
\end{align} 
\end{widetext}

\textbf{II. The case $c_0<0$}

Here we derive the dispersion relation for NSWs when $c_0<0$.
The method used here is exactly the same as in the previous case (for $c_0>0$). 
Using the analytical formula for the field profile in the nonlinear core [Eq.~(\ref{eqn:elip16f})] the field profiles in metal claddings given by Eqs.~(\ref{eqn:met_fields}), and Maxwell's equations [Eqs.~(\ref{eqn:max4})], we write the continuity conditions for the tangential electromagnetic field components $H_y$ and $E_z$ at the interfaces between the nonlinear core and the metal cladding $x=d$
The two resulting expressions for $H_d$ are compared and give the nonlinear dispersion relation in its final form for the case $c_0<0$
\begin{widetext}
\begin{align}
{k_0 q_3 \epsilon_{l,2}}\sqrt{A}\dn\left[ \sqrt{{\gamma^2}/{A}}  (d-x_0) \;   \big|\; m \right] = 
{ \epsilon_3 m } \sqrt{{ \gamma^2}}
\sn\left[ \sqrt{{{\gamma^2}/{A}}}\; (d-x_0) \;   \big|\; m \right]
\cn\left[ \sqrt{{{\gamma^2}/{A}}}\; (d-x_0) \;   \big|\; m \right].
\label{eqn:disp-JEM-2a}
\end{align}
\end{widetext}
Equations~(\ref{eqn:disp-JEM-1a}) and (\ref{eqn:disp-JEM-2a}) build the full dispersion relation for the NSW.
In order to obtain the dispersion diagram for a fixed structure and wavelength ($\epsilon_1$, $\epsilon_{l,2}$, $n_2^{(2)}$, $\epsilon_3$, $d$, $\lambda$) we scan $H_0$ values. 
For a fixed $H_0$, using Eq.~(\ref{eqn:c0}) we identify intervals of $\beta$ where $c_0$ is positive or negative. In these intervals, we find $\beta$ values that satisfy Eq.~(\ref{eqn:disp-JEM-1a}) or Eq.~(\ref{eqn:disp-JEM-2a}) depending on the sign of $c_0$.

\subsection{Interface model}
\label{sec:interface-model}

In Section~\ref{sec:JEM-theory}, we have derived the JEM that treats the Kerr nonlinearity present in the core of the NSW in a simplified way [see Eq.~(\ref{eqn:kerr_simple-jem})]. The used assumptions allowed us to obtain analytical formulas for the nonlinear dispersion relations and the field profiles of the nonlinear modes of the NSW. 

In this section, we will present the derivation of a model that is more numerical than the JEM but treats the Kerr-type nonlinearity in a more precise way. The permittivity of the nonlinear core in the frame of the interface model (IM) depends on all the components of the electric field and is described by
\begin{equation}
\epsilon_2(x) = \epsilon_{l,2} (x)+ \alpha_2 \left[E_x^2(x) + E_z^2(x)\right],
\label{eqn:kerr-im}
\end{equation} 
Moreover, there is no theoretical limitation of the values of the nonlinear permittivity change.

In the IM, the solutions of Maxwell's equations are sought numerically, as explained in the following. The field profiles inside the nonlinear core are found by numerical integration of Maxwell's equations that couple the $E_x$ and $E_z$ field components. The novelty of our numerical method lays in the fact that, the parameter space where the solutions are being sought is reduced by a constraint that is expressed in an analytical form. Below the derivation of this constraint  is presented and the numerical procedure of finding the nonlinear dispersion relations using the IM is shortly described.

\subsubsection{Analytical constraint}
\label{sec:anal-const}
  
The derivation of the IM starts from Maxwell's equations [Eqs.~(\ref{eqn:max4})]. In this approach the magnetic field is eliminated from these equations. The use of Eq.~(\ref{eqn:rotH41}) in Eqs.~(\ref{eqn:rotE4}) and (\ref{eqn:rotH42}) gives~\cite{Yin09,Walasik14,Walasik14a,Walasik-thesis14}
\begin{subequations}
 \begin{align}
\frac{\di  E_z}{\di x} &= k_0 \left( \beta - \frac{\epsilon}{\beta} \right)E_x,
\label{eqn:yin1}\\
\frac{\mathrm d  (\epsilon E_x)}{\di x} &=  \beta k_0  \epsilon E_z. 
\label{eqn:yin2}
\end{align} 
\end{subequations} 
In the nonlinear core Eqs.~(\ref{eqn:yin1}), (\ref{eqn:yin2}) and (\ref{eqn:kerr-im}) are transformed to (for the detailed derivation see Ref.~\cite{Walasik-thesis14})
 \begin{align}
\left( \frac{\di  E_z}{\di x} \right)^2 =   \left(\beta k_0\right)^2 E_x^2 & -  k_0^2 \epsilon_{l,2} \left( E_x ^2  +  E_z ^2 \right) \nonumber \\ 
 &-  k_0^2 \frac{\alpha_2}{2} \left(E_x^2 + E_z^2\right)^{2} + C_0,
\label{eqn:IM0}
\end{align} 
where the linear permittivity $\epsilon_{l,2}$, the nonlinear parameter $\alpha_2$ of the nonlinear core of the slot waveguide appear and $C_0$ denotes the integration constant.

In problems with a semi-infinite nonlinear medium the magnetic field $H_y$ and its $x$-derivative vanish when $x \rightarrow \pm \infty$. In such cases, these boundary conditions allow us to set the integration constant $C_0$ in Eq.~(\ref{eqn:IM0}) to zero~\cite{Walasik14}. Here we deal with a problem in which the nonlinear medium is sandwiched between two linear (metal) layers, and therefore the nonlinear medium has a finite size. In this case, the integration constant can not be set automatically to zero. 

We compare the right-hand side of Eq.~(\ref{eqn:IM0}) with the square of the right-hand side of Eq.~(\ref{eqn:yin1}). The comparison gives
\begin{align}
\left({\epsilon_2^2}/{\beta^2} - 2\epsilon_2 \right) E_z^2  +& \epsilon_{l,2} \left( E_x ^2  +  E_z ^2 \right) + \nonumber  \\ &{\alpha_2}/{2}\left(E_x^2 + E_z^2 \right)^2 =C_0.
\label{eqn:IM1}
\end{align} 
Equation~(\ref{eqn:IM1}) together with continuity conditions for the tangential components of the electromagnetic field will be helpful in finding the constraints reducing the parameter space where the solutions of Maxwell's equations are sought, in order to find the dispersion curves for the NSW configuration.

We use the continuity conditions for the tangential components of the electromagnetic field in order to relate the values of the electric field components $E_x$ and $E_z$ at the nonlinear interfaces to the values of the total electric field amplitude, defined as
\begin{equation}
E_p \equiv \sqrt{E_{x,p}^2 + E_{z,p}^2},
\label{eqn:total-elect}
\end{equation} 
where the additional subscript $p \in \{0,d\}$ denotes the $x$ coordinate at which the quantity is calculated.
The field distributions for the electric field components in the semi-infinite linear metal regions are found by solving linear wave equations for these components.
The components of the electric field in the cladding metal layers are given by:
\begin{enumerate}
\item
In the left metal region ($x<0$ --- layer 1 in Fig.~\ref{fig:geometry-slot}):
\begin{subequations}
\label{eqn:IM_Exz1}%
\begin{align}
E_{x,1} &= A_x e^{k_0 q_1 x} \label{eqn:IM_Ex1},\\
E_{z,1} &= A_z e^{k_0 q_1 x} \label{eqn:IM_Ez1}
\end{align}
\end{subequations}
\item
In the right metal region ($x>d$ --- layer 3 in Fig.~\ref{fig:geometry-slot}):
\begin{subequations}
\label{eqn:IM_Exz3}%
\begin{align}
E_{x,3} &= B_x e^{-k_0 q_3 (x-d)}  \label{eqn:IM_Ex3}, \\
E_{z,3} &= B_z e^{-k_0 q_3 (x-d)}  \label{eqn:IM_Ez3}. 
\end{align}
\end{subequations}
\end{enumerate}
Only one exponential term is present in each of the expressions so that the electric field decays exponentially for $x \rightarrow \pm \infty$.

Using the continuity conditions for the tangential field components ($H_y$ and $E_z$) at both interfaces ($x=0$ and $x=d$), and Eqs.~(\ref{eqn:rotH41}) and (\ref{eqn:rotH42}) in the linear layers, we can express the electric field components at the left interface ($E_{x,0}$, $E_{z,0}$) as a function of the total electric field amplitude at this interface $E_0$:
\begin{subequations}
\label{eqn:IM_E0_both}%
\begin{align}
E_{x,0}^2 &= {(\epsilon_{1} \beta E_0)^2}/[{(\epsilon_{2,0} q_1)^2 + (\epsilon_1 \beta)^2}], \label{eqn:IM_Ex0_fin}\\
E_{z,0}^2 &= {(\epsilon_{2,0} q_1 E_0)^2}/[{(\epsilon_{2,0} q_1)^2 + (\epsilon_1 \beta)^2}], \label{eqn:IM_Ez0_fin}
\end{align}
\end{subequations}
where $\epsilon_{2,0}$  denotes the value of the nonlinear permittivity at the left interface of the core and is equal to $\epsilon_2|_{x=0^+} = \epsilon_{l,2} + \alpha_2 E_0^2$.
Similarly, we can express the electric field components at the right interface ($E_{x,d}$, $E_{z,d}$) as a function of a total electric field amplitude at this interface $E_d$:
\begin{subequations}
\label{eqn:IM_Ed_both}%
\begin{align}
E_{x,d}^2 &= {(\epsilon_{3} \beta E_d)^2}[/{(\epsilon_{2,d} q_3)^2 + (\epsilon_3 \beta)^2}], \label{eqn:IM_Exd_fin}\\
E_{z,d}^2 &= {(\epsilon_{2,d} q_3 E_d)^2}/[{(\epsilon_{2,d} q_3)^2 + (\epsilon_3 \beta)^2}], \label{eqn:IM_Ezd_fin}
\end{align}
\end{subequations}
where $\epsilon_{2,d}$  denotes the value of the nonlinear permittivity at the right interface of the core and is equal to $\epsilon_2|_{x=d^-} = \epsilon_{l,2} + \alpha_2 E_d^2$.

Equation~(\ref{eqn:IM1}) can now be rewritten on each interface in such a way that, it depends only on the total electric field amplitude at this interface (as a parameter), the effective index $\beta$ as an unknown and the opto-geometric material parameters which are known and fixed for a given NSW. 
Inserting Eqs.~(\ref{eqn:IM_Ex0_fin}) and (\ref{eqn:IM_Ez0_fin}) into Eq.~(\ref{eqn:IM1}) taken at $x = 0^+$, we obtain the nonlinear dispersion relation at the left interface ($x= 0$):
\begin{equation}
 \frac{ \left[ \left( \frac{\epsilon_{2,0}}{\beta}\right)^2 - 2 \epsilon_{2,0}\right] (\epsilon_{1} \beta)^2}{(\epsilon_{2,0} q_1)^2 + (\epsilon_1 \beta)^2} + \epsilon_{l,2} + \frac{\alpha_2}{2}E_0^2  = \frac{C_0}{E_0^2}.
 \label{eqn:IM_disp0}
\end{equation}

Inserting Eqs.~(\ref{eqn:IM_Exd_fin}) and (\ref{eqn:IM_Ezd_fin}) into Eq.~(\ref{eqn:IM1})  taken at $x = d^-$, we obtain the dispersion relation at the right interface ($x=d$):
\begin{equation}
  \frac{ \left[ \left( \frac{\epsilon_{2,d}}{\beta}\right)^2 - 2 \epsilon_{2,d}\right] (\epsilon_{3} \beta)^2}{(\epsilon_{2,d} q_3)^2 + (\epsilon_3 \beta)^2} + \epsilon_{l,2} + \frac{\alpha_2}{2}E_d^2  = \frac{C_0}{E_d^2}.
 \label{eqn:IM_dispd}
\end{equation}

Comparing the dispersion relations for the single interfaces given by Eqs.~(\ref{eqn:IM_disp0}) and (\ref{eqn:IM_dispd}), we eliminate the integration constant $C_0$ and obtain the final equation of the IM: 
\begin{align}
\frac{ \left[ \left( \frac{\epsilon_{2,0}}{\beta}\right)^2 - 2 \epsilon_{2,0}\right]  \frac{(\epsilon_{1} \beta)^2}{(\epsilon_{2,0} q_1)^2 + (\epsilon_1 \beta)^2} + \epsilon_{l,2} + \frac{\alpha_2}{2}E_0^2 }
{ \left[ \left( \frac{\epsilon_{2,d}}{\beta}\right)^2 - 2 \epsilon_{2,d}\right] \frac{(\epsilon_{3} \beta)^2}{(\epsilon_{2,d} q_3)^2 + (\epsilon_3 \beta)^2} + \epsilon_{l,2} + \frac{\alpha_2}{2}E_d^2 }
= 
\frac{E_d^2}{ E_0^2}.
 \label{eqn:IM_final}
\end{align}
Equation~(\ref{eqn:IM_final}) represents the constraint that will be used in the numerical algorithm computing the dispersion curves presented in Section~\ref{sec:im-num-alg}, in order to reduce the dimension of the parameter space where the solutions of Maxwell's equations in NSW structures are sought.

It is worth noticing that, in Eq.~(\ref{eqn:IM_final}), neither the wavelength of light $\lambda$ nor the width of the waveguide core $d$ appear. This means that the condition given by Eq.~(\ref{eqn:IM_final}) is identical, regardless of the values of $\lambda$ and $d$. This condition depends only on the material parameters ($\epsilon_1$, $\epsilon_{l,2}$, $\alpha_2$, and $\epsilon_3$) and the field intensities at both nonlinear core interfaces ($E_0$ and $E_d$). The size of the core and the wavelength will appear in our next step --- the numerical integration of Maxwell's equations leading to the field profiles in the core of the waveguide and to the dispersion curves of the NSW. 

As stated before, Eq.~(\ref{eqn:IM_final}) is not a dispersion relation for the slot waveguide modes but only a constraint that limits the parameter space where the solutions in the frame of the IM can be found. In order to obtain the dispersion relations in the NSW, the field profiles in the core are found by numerical integration of Maxwell's equations. Maxwell's equations are written as a set of coupled equations relating both electric field components $E_x$ and $E_z$. These equations are derived from Eqs.~(\ref{eqn:yin1}) and (\ref{eqn:yin2}):
\begin{subequations}
\label{eqn:IM_field_all0}
\begin{align}
 \frac{\di  E_x}{\di x} &= k_0\frac{\beta  \epsilon_2  E_z - 2  \alpha_2  E_z  E_x^2  \left(\beta - \frac{\epsilon_2}{\beta}   \right)}{ \epsilon_2 + 2  \alpha_2  E_x^2},
\label{eqn:IM_field1} \\
\frac{\di  E_z}{\di x} &= k_0 \left( \beta - \frac{ \epsilon_2}{\beta} \right)E_x.
\label{eqn:IM_field2}
\end{align}
\end{subequations}

\subsubsection{Numerical algorithm and nonlinear dispersion relations}
\label{sec:im-num-alg}

In this section, we present the description of the numerical algorithm that is used to find the dispersion relation for the modes of a NSW in the frame of the IM. This algorithm uses the shooting method~\cite{Press07} to find solutions of Maxwell's equations in the waveguide core and finally the nonlinear dispersion relation for the NSW. 

In a general case, the numerical procedure would be the following. First, for a given structure, we fix the parameters $E_0$, $E_d$, and $\beta$ and integrate Maxwell's equations [Eqs.~(\ref{eqn:IM_field_all0})]  in the waveguide core with the values $E_0$ and $\beta$ as initial parameters. The results of the integration are the field profiles $E_x(x)$ and $E_z(x)$ inside the nonlinear core and, in particular, the computed total electric field amplitude at the interface $x=d$ denoted by $E_d^{\textrm{(num)}}$. Next, we verify if the result of the numerical integration fulfills the conditions resulting from the problem formulation:
\begin{enumerate}
\item \label{cond-1} Is $E_d^{\textrm{(num)}}$ equal to the initially fixed value $E_d$?
\item \label{cond-2} Does the $x$-derivative of the product of the permittivity and the transverse electric field $\epsilon_2 E_x$ at the interface $x=d^-$ have the correct sign? The condition for the correct sign reads
\begin{equation}
E_{x,d} \left. \frac{\mathrm d (\epsilon_2 E_x)}{\di x}\right|_{x=d^-}  = {- k_0 q_3 \epsilon_3 B_x^2}, 
\label{eqn:G} 
\end{equation}
and it is derived in Appendix~\ref{app:cond} and in Ref.~\cite{Walasik-thesis14}. 
The right-hand side of Eq.~(\ref{eqn:G}) is positive as $\epsilon_3$ of the metal is negative and all the other quantities there are positive. Therefore, the sign of the derivative $[\mathrm d (\epsilon_2 E_x)/\mathrm d x]|_{x=d^-}$ must be identical to the sign of $E_{x,d}$.
\item \label{cond-3} Do the components of electric field at $x=d$ fulfill the conditions given by Eqs.~(\ref{eqn:IM_Exd_fin}) and (\ref{eqn:IM_Ezd_fin})?
\end{enumerate}
Is these three conditions are fulfilled then the triplet ($E_0$, $E_d$, and $\beta$) and the corresponding field profiles are accepted as a genuine solution of our problem.

In the general case described above, the parameter space where the solutions are sought is three-dimensional and it is spanned by $E_0$, $E_d$, and $\beta$. However, we can separate the problem into two cases, where we will be able to simplify it and look for the solutions in only two-dimensional spaces. The two cases are :
\begin{enumerate}
\item
\label{case-sym}
The case of symmetric ($E_{x,0} = E_{x,d}$) or antisymmetric solutions ($E_{x,0} = -E_{x,d}$) (for both symmetric and antisymmetric solutions $E_{0} = E_{d}$) in a symmetric NSW ($\epsilon_1 = \epsilon_3$). In this case, we look for the solutions of Maxwell's equations in a two-dimensional space spanned by $E_0$ and $\beta$ for each of the cases $E_{x,0}  = \pm E_{x,d}$. For $E_{0} = E_{d}$ in symmetric structures, Eq.~(\ref{eqn:IM_final}) represents an identity and it is satisfied for all values of $\beta$. Therefore, Eq.~(\ref{eqn:IM_final}) will not provide any help in further reducing the parameter space where the solutions are sought.

\item
\label{case-asym}
The case of either the asymmetric solutions ($E_0  \neq E_d$) in a symmetric NSW structure ($\epsilon_1 = \epsilon_3$) or any solution in an asymmetric NSW ($\epsilon_1 \neq \epsilon_3$). For these types of solutions, Eq.~(\ref{eqn:IM_final}) is not an identity and results in a constraint on the three-dimensional space where the solutions are sought. 
Equation~(\ref{eqn:IM_final}) is transformed to the form:
\begin{equation}
p_4 \beta^4 + p_2 \beta^2 + p_0 = 0,
\label{eqn:IM_beta}
\end{equation}
where
\begin{widetext}
\begin{subequations}
\begin{align}
p_4 = \;& 2 \epsilon_{2,d} \epsilon_{3}^2 E_d^2 (\epsilon_{2,0}^2 + \epsilon_{1}^2) - 
      2 \epsilon_{2,0} \epsilon_{1}^2 E_0^2 (\epsilon_{2,d}^2 + \epsilon_{3}^2) +
      f (\epsilon_{2,0}^2 + \epsilon_{1}^2)(\epsilon_{2,d}^2 + \epsilon_{3}^2), \\
p_2 = \;& \epsilon_{2,0}^2 \epsilon_{1}^2 E_0^2 (\epsilon_{2,d}^2 + \epsilon_{3}^2) - 
      \epsilon_{2,d}^2 \epsilon_{3}^2 E_d^2 (\epsilon_{2,0}^2 + \epsilon_{1}^2)+  \nonumber \\ 
     & 2 \epsilon_{2,0} \epsilon_{2,d} \epsilon_{1} \epsilon_{3}
     (\epsilon_{2,d} \epsilon_{1} E_0^2 - \epsilon_{2,0} \epsilon_{3} E_d^2) - 
     f [\epsilon_{2,0}^2 \epsilon_{1} (\epsilon_{2,d}^2 + \epsilon_{3}^2) + 
        \epsilon_{2,d}^2 \epsilon_{3} (\epsilon_{2,0}^2 + \epsilon_{1}^2)], \\
p_0 = \;& \epsilon_{2,0}^2 \epsilon_{2,d}^2 \epsilon_{1} \epsilon_{3} 
        (\epsilon_{3} E_d^2 - \epsilon_{1} E_0^2 + f), \\
f = \;& \epsilon_{l,2}(E_0^2 + E_d^2) + \frac{\alpha_2}{2} (E_0^4 + E_d^4).
\end{align}
\end{subequations}
\end{widetext}

This shows that Eq.~(\ref{eqn:IM_final}) is satisfied only by a finite set of $\beta$ values.
Because we look for forward propagating nonlinear modes and material losses are neglected, the only physically meaningful solutions of Eq.~(\ref{eqn:IM_beta}) are the ones where $\beta$ is real and positive. Therefore, the two possible roots are:
\begin{equation}
\beta_{\pm} = \sqrt{\frac{-p_2 \pm\sqrt{p_2^2 - 4 p_4 p_0}}{2 p_4}},
\label{eqn:beta-reduce}
\end{equation}
Only the real solutions among $\beta_{\pm}$ are used further in the process of the resolution of the nonlinear problem.

Using Eq.~(\ref{eqn:IM_final}), the three-dimensional space from the general case is now reduced to a two-dimensional space [one for each of the real effective indices given by Eq.~(\ref{eqn:beta-reduce})] spanned by $E_0$ and $E_d$. 
Therefore, instead of scanning a full three-dimensional space spanned by $E_0$ and $E_d$, and $\beta$ we need to scan only a few two-dimensional spaces spanned by $E_0$ and $E_d$ corresponding to  the physically meaningful $\beta$ values given by Eq.~(\ref{eqn:beta-reduce}). In other words, for a pair ($E_0$, $E_d$) we just need to check if the field integration gives valid results (i.e., if conditions~\ref{cond-1}--\ref{cond-3} are fulfilled) for real values among $\beta_{\pm}$. If all the conditions are fulfilled then the triplet ($E_0$, $E_d$, and $\beta$) and the corresponding field profiles are accepted as a genuine solution of our problem.
\end{enumerate}

\section{Limiting cases for two-layer structures}
\label{sec:ver}

In this section we will derive the expressions for the dispersion relations in the limiting case of the single interface between a metal and a nonlinear dielectric. This will prove that our models reproduce already known results for simpler structures. This limiting case dispersion relations will also provide approximated analytical expressions for the propagation constant of highly asymmetric modes that resemble nonlinear plasmons on a single interface as shown in the following article~\cite{Walasik14c}.

\subsection{Jacobi elliptic function based model}
\label{sec:ver-fbm}

In Sec.~\ref{sec:JEM-NL-field-prof}, we have stated that in case of a semi-infinite nonlinear medium (a single interface between a metal and a nonlinear dielectric) the integration constant in Eq.~(\ref{eqn:first_int}) should be set to zero as both the magnetic field $H_y$ and its derivative tend to zero at infinity.  One way to find the dispersion relation for nonlinear waves propagating along a single interface is to use Eq.~(\ref{eqn:c0}). Setting $c_0 = 0$ in this equation we obtain an analytical formula for the effective indices
\begin{equation}
 \left(  {\epsilon_{l,2}}/{\epsilon_1}\right)^2 q_1^2 - q_2^2 + \frac{a_2}{2} H_0^2     = 0.
\label{eqn:c0approx}
\end{equation} 
Using the definitions of $q_k$ and $a_2$ [see Eqs.~(\ref{eqn:q-def}) and (\ref{eqn:a-def})] in  Eq.~(\ref{eqn:c0approx}), we find the analytical expression for the effective index of a nonlinear wave at a single interface between a metal and a nonlinear dielectric in an explicit form:
\begin{equation}
\beta = \sqrt{\frac{\epsilon_1\epsilon_{l,2}(\epsilon_{l,2}-\epsilon_1)}{\epsilon_{l,2}^2-\epsilon_1^2 + \frac{n_2^{(2)}\epsilon_1^2 H_0^2}{2 \epsilon_0 c \epsilon_{l,2}}}}.
\label{eqn:beta_approx}
\end{equation} 
Another way to find the dispersion relation for nonlinear waves propagating along a single interface is to set $c_0$ at the later stage of the JEM derivation, namely in Eqs.~(\ref{eqn:disp-JEM-1a}) or (\ref{eqn:disp-JEM-2a}) and the related parameters [Eqs.~(\ref{eqn:gam_del1a}) or (\ref{eqn:gam_del2}), respectively]. In both cases this procedure yields~\cite{Walasik-thesis14} (using the expressions for the limiting values of Jacobi elliptic functions in the case of the parameter $m=1$ provided in Ref.~\cite{Abramowitz72})
\begin{align}
\tanh \left[ k_0 q_2 (d - x_0) \right] = \frac{\epsilon_{l,2} q_3}{\epsilon_3 q_2}.
\label{eqn:lim-disp-I2}
\end{align} 
Equation~(\ref{eqn:lim-disp-I2}) describes the nonlinear dispersion relation for plasmon--solitons on a single interface (at $x=d$) between a metal and a nonlinear dielectric. Equation~(\ref{eqn:lim-disp-I2}) is equivalent to Eq.~(42) from Ref.~\cite{Walasik14}, which gives the dispersion relation for a single metal/nonlinear dielectric interface obtained using the field based model developed in Ref.~\cite{Walasik14}, taking into account that: (i) in the frame of the JEM we used the assumption that $\epsilon_2 = \epsilon_{l,2}$ in the continuity conditions used to derive the nonlinear dispersion relations (see Section~\ref{sec:disp-rel-slot}, and (ii) in Ref.~\cite{Walasik14} the interface is located at $x=0$.

Equations~(\ref{eqn:beta_approx}) and (\ref{eqn:lim-disp-I2}) give also approximated expressions for the effective indices of highly asymmetric solutions in NSWs, as it will be proven by the numerical results presented in Ref.~\cite{Walasik14c}. Highly asymmetric solutions are strongly localized on one of the interfaces and therefore the problem can be simplified to a single-interface problem. A comparison of the approximated solution given by Eqs.~(\ref{eqn:beta_approx}) and (\ref{eqn:lim-disp-I2}) with the exact solutions of the JEM will be given in Sec.~\ref{sec:res-sym} of Ref.~\cite{Walasik14c}.

It is worth noting that using Eq.~(\ref{eqn:beta_approx}) in the linear case ($H_0 \rightarrow 0$ or $n_2^{(2)} \rightarrow 0$), we recover the dispersion relation for a linear surface plasmon propagating along a single interface \{see Eq.~(2.14) in Ref.~\cite{Maier07}\}
\begin{equation}
\beta = \sqrt{{\epsilon_1\epsilon_{l,2}}/({\epsilon_1 + \epsilon_{l,2} })}.
\label{eqn:beta_approx_2}
\end{equation} 

\subsection{Interface model}

Equations~(\ref{eqn:IM_disp0}) and (\ref{eqn:IM_dispd}) considered separately give the dispersion relation for a single interface between a metal and a nonlinear dielectric. In this case, the nonlinear medium is semi-infinite which means that $C_0$ must be set to zero (see discussion in Sec.~\ref{sec:anal-const}). Setting $C_0 = 0$ in Eq.~(\ref{eqn:IM_disp0}) yields  
\begin{equation}
 \frac{  \left[ \left( \frac{\epsilon_{2,0}}{\beta}\right)^2 - 2 \epsilon_{2,0}\right] (\epsilon_{1} \beta)^2}{(\epsilon_{2,0} q_1)^2 + (\epsilon_1 \beta)^2} + \epsilon_{l,2} + \frac{\alpha_2}{2}E_0^2  = 0.
 \label{eqn:IM_disp-single}
\end{equation}
This equation can be solved analytically for $\beta$. The solution depends on the parameters of the structure ($\epsilon_1$, $\epsilon_{l,2}$, $\alpha_2$, $\epsilon_3$) and the electric field amplitude at the interface $E_0$ and is given by 
\begin{equation}
\beta = \sqrt{\frac{\epsilon_1 \epsilon_{2,0}^2 (\epsilon_{l,2} - \epsilon_1 + \frac{\alpha_2}{2} E_0^2)}{(\epsilon_{2,0}^2 + \epsilon_1^2)(\epsilon_{l,2} + \frac{\alpha_2}{2} E_0^2) - 2\epsilon_1^2 \epsilon_{2,0}}}.
 \label{eqn:IM_single-anal}
\end{equation}
Only the positive root is considered because we are interested here in forward propagating waves only.
Equation~(\ref{eqn:IM_single-anal}) can be compared to Eq.~(11) in Ref.~\cite{Huang09} and to Eq.~(14) in Ref.~\cite{Yin09} derived for the case of a single metal/nonlinear dielectric interface.

The solution for a single-interface problem provides a good approximation (in terms of effective index $\beta$ and the field profiles) for first-order highly asymmetric modes in the NSWs whose field profiles are mostly localized on one interface of the core (see also the discussion in Sec.~\ref{sec:ver-fbm}). These solutions are invariant with respect to the waveguide width as they interact strongly only with one of the core interfaces. More comments and illustrations of this property will be presented in Section~\ref{sec:results-single-interface-limit} in Ref.~\cite{Walasik14c}, where we discuss the results for the symmetric NSWs. In the linear limit $\alpha_2 E_0^2 \rightarrow 0$, Eq.~(\ref{eqn:IM_single-anal}) transforms to Eq.~(\ref{eqn:beta_approx_2}), as expected.

\section{Conclusions}

We have presented two complementary models based on Maxwell's equations to study the properties of the stationary
TM solutions in planar nonlinear structures made of a focusing Kerr nonlinear dielectric core 
surrounded by semi-infinite metal regions.
The first model uses a simplified treatment of nonlinearity that takes into account only the transverse component of the electric field in the nonlinear term and assumes that the nonlinear refractive index change is small compared to the  linear refractive index. 
It provides an analytical description of both the field profiles in the whole waveguide using Jacobi elliptic special functions and the nonlinear dispersion relations.  
These features allow one to study rapidly and accurately the properties of the NSW structures as a function of their opto-geometric parameters.
The second model  takes into account the full dependency of the Kerr nonlinear term on all electric field components and no assumption is required on the amplitude of the nonlinear term. It allows us both to determine the domain of validity of the first model and to investigate with accuracy the effects of high nonlinearities.
These two models prove their usefulness in the next article (Ref.~\cite{Walasik14c}) where we provide a complete and accurate study of nonlinear slot waveguides. It is worth mentionning that these models can be extented to more complicated  geometries and refractive index forms.

\begin{acknowledgments}
This work was supported by the European Commission through the Erasmus Mundus Joint Doctorate Programme  Europhotonics (Grant No. 159224-1-2009-1-FR-ERA MUNDUS-EMJD). This work was not funded by the French Agence Nationale de la Recherche.
\end{acknowledgments}

\vspace{0.25cm}

\appendix

\section{JEM}
\label{app:jem}
In this appendix we present the derivation of the solution of Eq.~(\ref{eqn:first_int}) that leads to the expressions for the field profiles in the nonlinear core in the frame of the Jacobi elliptic function based model (see Sec.~\ref{sec:JEM-NL-field-prof}). The derivation is presented in separated cases depending on the signs of the integration constant $c_0$ and of the $q_2^2$ parameter. This separation allows us to work with positive quantities only in each of the cases, which facilitates the choice of ambiguous signs.

In this appendix we first divide the problem according to the sign on $q_2^2$. For each of the possible signs we divide the problem according to the $c_0$ sign. This choice is dictated by the fact that in the case of negative $q_2^2$ we have to consider only the case of positive $c_0$ as it can be seen from Eq.~(\ref{eqn:c0}).

\subsection{The case $q_2^2>0$}
\label{sec:1}
\subsubsection{The subcase $c_0>0$}
\label{sec:1a}

At first, we consider the case where $q_2^2>0$ and $c_0>0$ and find the solutions of the nonlinear wave equation [Eq.~(\ref{eqn:first_int})] for this case. 
Equation~(\ref{eqn:first_int}) was transformed into Eq.~(\ref{eqn:first_3}) and its left-hand side can be expressed in the form of the integrand of an elliptic integral (see Ref.~\cite{Abramowitz72}):
\begin{equation}
\frac{\di H_2}{\sqrt{(\gamma^2 + H_2^2)(\delta^2 -H_2^2)}} = \pm\; \sqrt{\frac{1}{A}}\; \di x,
\label{eqn:elip1-app}
\end{equation} 
where the parameters $\gamma$ and $\delta$ were introduced. In order to relate the newly introduced parameters with parameters $A$, $Q$, and $c_0$, we compare the expressions under the square-root on the left-hand sides of Eqs.~(\ref{eqn:first_3}) and (\ref{eqn:elip1-app}). This comparison results in 
\begin{subequations} 
\label{eqn:gam_del1a-app}
\begin{align}
\gamma^2 &= ({ \sqrt{A^2Q^2 + 4 A c_0} - AQ})/{2}, \label{eqn:gam2-app} \\
\delta^2 & =  ({ \sqrt{A^2Q^2 + 4 A c_0} + AQ})/{2}, \label{eqn:del2-app}
\end{align}
\end{subequations} 
where the choice of the sign in front of the $AQ$ term is dictated by our assumption that the magnetic field $H_y$ is real [see the assumption made in Eq.~(\ref{eqn:harmonic1}) and Ref.~\cite{Walasik-thesis14} for more details]. 

Integrating Eq.~(\ref{eqn:elip1-app}), we obtain
\begin{equation}
\int_{H_2(0)}^{H_2(x)} \frac{\di H_2}{\sqrt{(\gamma^2 + H_2^2)(\delta^2 -H_2^2)}} = \pm\; \sqrt{\frac{1}{A}}\; x.
\label{eqn:elip2-app}
\end{equation} 
The integral on the left-hand side of Eq.~(\ref{eqn:elip2-app}) can be separated into two integrals:
\begin{align}
\int_{H_2(0)}^{H_2(x)}  =
\int_{H_2(0)}^{\delta}  - 
\int_{H_2 (x)}^{\delta}.
\label{eqn:elip3-app}
\end{align}
Using this fact, Eq.~(\ref{eqn:elip2-app}) yields
\begin{widetext}
\begin{equation}
\int_{H_2(0)}^{\delta} \frac{\di  H_2}{\sqrt{(\gamma^2 + H_2^2)(\delta^2 -H_2^2)}} - \int_{H_2 (x)}^{\delta} \frac{\di H_2}{\sqrt{(\gamma^2 + H_2^2)(\delta^2 -H_2^2)}} = \pm\; \sqrt{\frac{1}{A}}\; x.
\label{eqn:elip4-app}
\end{equation} 
\end{widetext}
Multiplying Eq.~(\ref{eqn:elip4-app}) by $\sqrt{\gamma^2+\delta^2}$ and using formula 17.4.52 from Ref.~\cite{Abramowitz72}, gives
\begin{equation}
\cn^{-1}\left[\frac{H_2(0)}{\delta} \bigg| m\right] - \cn^{-1}\left[\frac{H_2(x) }{ \delta}  \bigg|m\right] = \pm\; \sqrt{{\frac{s}{A}}}\; x,
\label{eqn:elip5-app}
\end{equation} 
where $\cn^{-1}(u|m)$ is the inverse of the Jacobi elliptic function $\cn(u|m)$ and 
\begin{subequations}
\label{eqn:params1-app}
\begin{align}
m &= {\delta^2}/{s}, \\
s &= \gamma^2 + \delta^2.
\end{align}
\end{subequations}
Jacobi elliptic functions are defined with the argument $u$ and the parameter $m$~\cite{Abramowitz72}.
Reorganizing the terms and applying the Jacobi elliptic function $\cn$ to both sides of Eq.~(\ref{eqn:elip5-app}), results in the expression for the magnetic field in the core of a NSW in the case where both $c_0$ and $q_2^2$ are positive:
\begin{align}
H_2(x) &=  \delta \cn\left\{  \mp\; \sqrt{{{s}/{A}}}\; (x- x_0)   \big| m \right\},
\label{eqn:elip6-app}\\
x_0 &= -\sqrt{{{A}/{s}}} \cn^{-1}\left[{H_2(0)}/{\delta} \big| m\right]. \label{eqn:x01-app}
\end{align} 

The uncertainty about the sign in front of the square-root in the argument of the $\cn$ function in Eq.~(\ref{eqn:elip6-app}) can be resolved by the analysis of Eq.~(\ref{eqn:elip2-app}) using the properties of the Jacobi elliptic function $\cn$ and the field continuity conditions at the interface $x=0$. Without loss of generality of the obtained results, we assume that in $H_0>0$ and $d>0$. In this case, the proper choice of the sign in front of the square-root in Eq.~(\ref{eqn:elip6-app}) is the bottom plus sign (see Ref.~\cite{Walasik-thesis14} for more details).
Finally, the field profile in the nonlinear core for the subcase I.a is given by
\begin{equation}
H_2(x) =  \delta \cn\left\{   \sqrt{{{s}/{A}}}\; (x-x_0)  \big| m \right\}.
\label{eqn:elip6f-app}
\end{equation}

\subsubsection{The subcase $c_0<0$}

Here we consider the case where $q_2>0$ and $c_0<0$. In order to work with positive quantities only, we substitute $c_0$ by $-|c_0|$ in Eq.~(\ref{eqn:first_3}) and obtain
\begin{equation}
\frac{\di H_2}{\sqrt{-A |c_0| + A Q H_2^2 - H_2^4}} = \pm\; \sqrt{\frac{1}{A}}\; \di x.
\label{eqn:first_4-app}
\end{equation}  
The left-hand side of Eq.~(\ref{eqn:first_4-app}) is then expressed in the form of the integrand of an elliptic integral (see Ref.~\cite{Abramowitz72}):
\begin{equation}
\frac{\di H_2}{\sqrt{(\gamma^2 - H_2^2)(H_2^2 - \delta^2 )}} = \pm\; \sqrt{\frac{1}{A}}\; \di x.
\label{eqn:elip10-app}
\end{equation} 

Comparing the expressions under the square-root on the left-hand sides of Eqs.~(\ref{eqn:first_4-app}) and (\ref{eqn:elip10-app}) allows us to find the relations between the new parameters $\gamma$ and $\delta$ and the parameters~$A$,~$Q$,~and~$c_0$:
\begin{subequations} 
\label{eqn:gam_del2-app}
\begin{align}
\gamma^2 = ({ AQ + \sqrt{A^2Q^2 - 4 A |c_0|}})/{2}, \\
\delta^2 =  ({ AQ - \sqrt{A^2Q^2 - 4 A |c_0|}})/{2}.
\end{align}
\end{subequations} 
The choice of the signs in front of the square-roots is Eqs.~(\ref{eqn:gam_del2-app}) is arbitrary due to the symmetric role of $\gamma$ and $\delta$ in Eq.~(\ref{eqn:elip10-app}) and due to the fact that $\delta^2 = AQ - \gamma^2$.

The expression that appears under the square-root in Eqs.~(\ref{eqn:gam_del2-app}) is a difference of two positive quantities. In order for $\gamma$ and $\delta$ to be real (which ensures that the magnetic field $H_2$ is real), the quantity under the square-root must be positive or equal to zero. Writing expression under the square-root explicitly using the definitions of $A$, $Q$ [Eqs.~(\ref{eqn:def-AQ-finite})], and $c_0$ [Eq.~(\ref{eqn:c0})] we obtain
\begin{align}
A^2 Q^2 &+ 4 A c_0 = \nonumber \\ &\left( q_2^2 - a_2 H_0^2 \right)^2 + 2 a_2 H_0^2 \left( {\epsilon_2|_{x=0^+}}/{\epsilon_1} \right)^2 q_1^2,
\end{align} 
which is greater or equal to zero because both terms in the sum are greater or equal to zero. This proves that $\gamma$ and $\delta$ are real quantities.

Having found the expressions for $\gamma$ and $\delta$, we proceed with the derivation process. Integrating Eq.~(\ref{eqn:elip10-app}) gives 
\begin{equation}
\int_{H_2(0)}^{H_2(x)} \frac{\di H_2}{\sqrt{(\gamma^2 - H_2^2)(H_2^2 - \delta^2 )}} = \pm\; \sqrt{\frac{1}{A}}\; x.
\label{eqn:elip12-app}
\end{equation} 
This time, the integral on the left-hand side of Eq.~(\ref{eqn:elip12-app}) is separated in the following way:
\begin{align}
\int_{H_2(0)}^{H_2(x)}  =
\int_{H_2(0)}^{\gamma} -
\int_{H_2 (x)}^{\gamma} .
\label{eqn:elip3b-app}
\end{align}
Inserting Eq.~(\ref{eqn:elip3b-app}) into Eq.~(\ref{eqn:elip12-app}), we obtain
\begin{widetext}
\begin{equation}
\int_{H_2(0)}^{\gamma}\frac{\di H_2}{\sqrt{(\gamma^2 - H_2^2)(H_2^2 - \delta^2 )}} - \int_{H_2 (x)}^{\gamma} \frac{\di H_2}{\sqrt{(\gamma^2 - H_2^2)(H_2^2 - \delta^2 )}} = \pm\; \sqrt{\frac{1}{A}}\; x.
\label{eqn:elip14-app}
\end{equation} 
\end{widetext}
Multiplying Eq.~(\ref{eqn:elip14-app}) by $\gamma$ and using formula 17.4.44 from Ref.~\cite{Abramowitz72} yields
\begin{equation}
\dn^{-1}\left[\frac{H_2(0)}{ \gamma} \bigg| m \right] - \dn^{-1}\left[\frac{H_2(x)}{ \gamma}  \bigg| m\right]  = \pm\; \sqrt{{\frac{\gamma^2}{A}}}\; x,
\label{eqn:elip15-app}
\end{equation} 
where $\dn^{-1}$ is the inverse of the Jacobi elliptic function $\dn$ and
\begin{align}
m &= ({\gamma^2 - \delta^2})/{\gamma^2}. \label{eqn:param2-app}
\end{align}
Reorganizing the terms and applying the Jacobi elliptic function $\dn$ to both sides of Eq.~(\ref{eqn:elip15-app}) results in the expression for the magnetic field in the core of a NSW for positive $q_2^2$ and negative $c_0$:
\begin{align}
H_2(x) &=  \gamma \dn\left\{ \sqrt{{{\gamma^2}/{A}}}\; (x -x_0)   \big| m \right\},
\label{eqn:elip16f-app}\\
x_0 &= - \sqrt{{{A}/{\gamma^2}}}\dn^{-1}\left[{H_2(0)}/{ \gamma} \big| m\right]. \label{eqn:x02-app}
\end{align} 
where the sign in front of the square root is chosen in a similar manner as in the case $c_0>0$.

\subsection{The case $q_2^2<0$}
\label{sec:2}

Here we consider the case of $q_2^2<0$. As stated in Sec.~\ref{sec:JEM-theory} below Eq.~(\ref{eqn:c0}), in this case, the integration constant $c_0$ takes only positive values. In the following, we find solutions of the nonlinear wave equation [Eq.~(\ref{eqn:wave5})] for this case. 

In order to work only with positive quantities in Eq.~(\ref{eqn:first_2}), in the case of negative $q_2^2$, we will substitute $q_2^2$ by its negative absolute value $-|q_2^2|$. This substitution transforms  Eq.~(\ref{eqn:first_2}) into
\begin{equation}
\frac{\di H_2}{\sqrt{c_0 - k_0^2 |q_2^2| H_2^2 - k_0^2 \frac{a_2}{2}H_2^4}} = \pm \di x.
\label{eqn:first_2_2-app}
\end{equation}  
We redefine $Q$ given by Eq.~(\ref{eqn:def-Q-finite}) to be positive. In the case II $Q$ is defined by 
\begin{align}
Q = k_0|q_2^2|
\label{eqn:q-II-app}
\end{align}
and $A$ is still defined by Eq.~(\ref{eqn:def-A-finite}). Using this definition, Eq.~(\ref{eqn:first_2_2-app}) can be written in the form:
\begin{equation}
\frac{\di H_2}{\sqrt{A c_0 - A Q H_2^2 - H_2^4}} = \pm\; \sqrt{\frac{1}{A}}\; \di x.
\label{eqn:first_3_2-app}
\end{equation}  
We rewrite Eq.~(\ref{eqn:first_3_2-app}) in the form of the elliptic integral:
\begin{equation}
\frac{\di H_2}{\sqrt{(\gamma^2 + H_2^2)(\delta^2 -H_2^2)}} = \pm\; \sqrt{\frac{1}{A}}\; \di x,
\label{eqn:elip1_2-app}
\end{equation} 
Comparing the expressions under the square-root on left-hand sides of Eqs.~(\ref{eqn:first_3_2-app}) and (\ref{eqn:elip1_2-app}) allows us to find the relations between the new parameters $\gamma$ and $\delta$ and the parameters~$A$,~$Q$,~and~$c_0$:
\begin{subequations} 
\label{eqn:gam_del1a_2-app}
\begin{align}
\gamma^2 &= ({ \sqrt{A^2Q^2 + 4 A c_0} + AQ})/{2} \label{eqn:gam4}, \\
\delta^2 & =  ({ \sqrt{A^2Q^2 + 4 A c_0} - AQ})/{2}. \label{eqn:del4}
\end{align}
\end{subequations} 

The following derivation is exactly the same as in the case I.a but we have to keep in mind that the definitions of $\gamma$ and $\delta$ are reversed [compare Eqs.~(\ref{eqn:gam_del1a-app}) and (\ref{eqn:gam_del1a_2-app})] and that $Q$ is defined now by Eq.~(\ref{eqn:q-II-app}).
Equation~(\ref{eqn:elip1_2-app}) is integrated and yields a formula that is identical to Eq.~(\ref{eqn:elip2-app}):
\begin{equation}
\int_{H_2(0)}^{H_2(x)} \frac{\di H_2}{\sqrt{(\gamma^2 + H_2^2)(\delta^2 -H_y^2)}} = \pm\; \sqrt{\frac{1}{A}}\; x.
\label{eqn:elip2_2-app}
\end{equation} 
By an analogy to the case presented in Sec.~\ref{sec:1a}, the final expression for the field profile in the nonlinear core for $q_2^2<0$ has  the form:
\begin{align}
H_2(x) &=  \delta \cn\left\{   \sqrt{{{s}/{A}}}\; (x-x_0)  \big| m. \right\},
\label{eqn:elip6f_2-app}\\
x_0 &= -\sqrt{{{A}/{s}}} \cn^{-1}\left[{H_2(0)}/{\delta} \big| m\right].
\end{align}

\subsection[Summary of the field profiles in the nonlinear core]{Summary and unification of the expressions for field profiles in the nonlinear core}
\label{sec:field-slot-summary}

In Sec.~\ref{sec:JEM-NL-field-prof} we have split the problem into two cases depending on the sign of the integration constant $c_0$ instead of three cases presented in this appendix. Below we show how to obtain the splitting presented in Sec.~\ref{sec:JEM-NL-field-prof} from the results derived in Appendices~\ref{sec:1} and~\ref{sec:2}.

We notice that the case presented in Secs.~\ref{sec:1a} and \ref{sec:2} can be merged into one case.
The sole difference between the expressions obtained in the case \ref{sec:1a} and \ref{sec:2} is in the formulas for the parameters $\gamma$ and $\delta$ [compare Eqs.(\ref{eqn:gam_del1a-app}) and Eqs.(\ref{eqn:gam_del1a_2-app})]. The sign in front of $Q$ is reversed. This difference is caused by our choice of the definition of the $Q$ parameter  [compare Eqs~(\ref{eqn:def-Q-finite}) and (\ref{eqn:q-II-app})]. This choice allowed us to work with positive quantities only which simplified the procedure of determination of the ambiguous signs. During the derivation of the nonlinear dispersion relations we will not encounter such problems any more and we can work with $Q$ that is either positive or negative.  
Therefore, without loss of generality we unify the cases presented in Secs.~\ref{sec:1a} and \ref{sec:2} to a common case where $c_0>0$. Now the distinction between the cases will be based on the sign of the integration constant $c_0$.

In the derivation of the dispersion relations presented in Sec.~\ref{sec:disp-rel-slot} we have considered two cases: I ($c_0>0$) for which we will use Eqs.~(\ref{eqn:def-AQ-finite}), (\ref{eqn:gam_del1a-app}), (\ref{eqn:params1-app}), (\ref{eqn:x01-app}), and (\ref{eqn:elip6f-app}) [presented as Eqs.~(\ref{eqn:elip6}) and (\ref{eqn:gam_del1a}) in Sec.~\ref{sec:JEM-NL-field-prof}]; and II ($c_0<0$) for which Eqs.~ (\ref{eqn:def-AQ-finite}), (\ref{eqn:gam_del2-app}), (\ref{eqn:param2-app})--(\ref{eqn:x02-app}) [presented as Eqs.~(\ref{eqn:elip16f}) and (\ref{eqn:gam_del2}) in Sec.~\ref{sec:JEM-NL-field-prof}] will be used.

\section{Derivation for the analytical condition in the numerical algorithm}
\label{app:cond}

In this appendix we present the derivation of the condition \ref{cond-2} in Sec.~\ref{sec:im-num-alg} used in the numerical algorithm of the IM. 
We start the derivation from the continuity condition for the $z$-component of the electric field at the interface $x=d$:
\begin{equation}
E_{z,2}|_{x=d^-} = E_{z,3}|_{x=d^+}.
\label{eqn:A} 
\end{equation} 
The longitudinal field component $E_z$ can be expressed with the help of Eq.~(\ref{eqn:yin2}) as
\begin{equation}
E_z = \frac{1}{k_0 \beta \epsilon}\frac{\mathrm d  (\epsilon E_x)}{\di x},
\label{eqn:B} 
\end{equation}
where we consider isotropic medium for which $\epsilon_x = \epsilon_z = \epsilon$. In the case of the uniform linear medium (as it is the case for layer 3), Eq.~(\ref{eqn:B}) simplifies to
\begin{equation}
E_z = \frac{1}{k_0 \beta}\frac{\mathrm d  E_x}{\di x}.
\label{eqn:C} 
\end{equation}
In the nonlinear medium (layer 2) Eq.~(\ref{eqn:B}) is written in the form
\begin{equation}
E_z = \frac{1}{k_0 \beta \epsilon_2}\frac{\mathrm d (\epsilon_2 E_x)}{\di x}.
\label{eqn:D} 
\end{equation}
Inserting Eqs.~(\ref{eqn:C}) and (\ref{eqn:D}) into Eq.~(\ref{eqn:A}) yields
\begin{equation}
\frac{1}{\epsilon_{2,d}} \left. \frac{\mathrm d (\epsilon_2 E_x)}{\di x}\right|_{x=d^-}  = \left. \frac{\di  E_x}{\di x} \right|_{x=d^+}.
\label{eqn:E} 
\end{equation}
Next we use the continuity condition for $H_y$ at the interface $x=d$
\begin{equation}
H_{y,2}|_{x=d^-} = H_{y,3}|_{x=d^+},
\end{equation}
which with the help of Eqs.~(\ref{eqn:rotH41}) and (\ref{eqn:IM_Ex3}) becomes
\begin{equation}
\epsilon_{2,d} E_{x,d} = \epsilon_3 B_x.
\label{eqn:deriv}
\end{equation}
Equation~(\ref{eqn:E}) is now multiplied by Eq.~(\ref{eqn:deriv}) to give
\begin{equation}
E_{x,d}\left. \frac{\mathrm d (\epsilon_2 E_x)}{\di x}\right|_{x=d^-}  = \epsilon_3 B_x \left. \frac{\di  E_x}{\di x} \right|_{x=d^+}.
\label{eqn:F} 
\end{equation}
Computing the derivative on the right-hand side of Eq.~(\ref{eqn:F}) using Eq.~(\ref{eqn:IM_Ex3}) gives [see Eq.~(\ref{eqn:G})]
\begin{equation}
E_{x,d} \left. \frac{\mathrm d (\epsilon_2 E_x)}{\di x}\right|_{x=d^-}  = {- k_0 q_3 \epsilon_3 B_x^2}.
\label{eqn:G-app} 
\end{equation}

\bibliographystyle{apsrev4-1}
\bibliography{PRA-14-part1-v3}{}

\end{document}